\documentclass[pdftex,citeautoscript,aps,prb,superscriptaddress,cite,amsmath,amsfonts,amssymb,twocolumn,showpacs]{revtex4}
\renewcommand{\vec}[1]{\mathbf{#1}} 
\usepackage{color}
\usepackage{wasysym}
\usepackage[pdftex]{graphicx}

\newcommand{\figref}[1]{Fig.~\ref{fig:#1}}

\newcommand{\eqnumref}[1]{(\ref{eq:#1})}
\renewcommand{\eqref}[1]{Eq.~\eqnumref{#1}}

\newcommand{\citeasnoun}[1]{Ref.$\;$\onlinecite{#1}}
\newcommand{\secref}[1]{Sec.~\ref{sec:#1}}

\definecolor{BrickRed}{cmyk}{0,0.89,0.94,0.28}
\definecolor{MidnightBlue}{cmyk}{0.98,0.13,0,0.43}
\definecolor{DarkGreen}{rgb}{0,0.7,0.1}

\def\a{s}
\def\b{s}

\newcommand{\add}[1]{\if\a\b{{\color{magenta} #1}}\else{#1}\fi}
\newcommand{\comm}[1]{\if\a\b{{\color{blue}\{\small \sc #1\}}}\else{}\fi}
\newcommand{\del}[1]{{\if\a\b{{\color{DarkGreen}[[#1]]}}\else{}\fi}}

\begin{document}

\title{Computation and visualization of photonic quasicrystal spectra
via Bloch's theorem}

\author{Alejandro W. Rodriguez}
\email{alexrod7@mit.edu}
\affiliation{Department of Physics, Massachusetts Institute of Technology, Cambridge, MA 02139}
\author{Alexander P. McCauley}
\thanks{A.M and A.R. contributed equally to this work}
\affiliation{Department of Physics, Massachusetts Institute of Technology, Cambridge, MA 02139}
\author{Yehuda Avniel}
\affiliation{Research Laboratory of Electronics, Department of Electrical Engineering and Computer Science, Massachusetts Institute of Technology, Cambridge, MA 02139}
\author{Steven G. Johnson}
\affiliation{Department of Mathematics,
Massachusetts Institute of Technology, Cambridge, MA 02139}

\begin{abstract}
Previous methods for determining photonic quasicrystal (PQC) spectra
have relied on the use of large supercells to compute the
eigenfrequencies and/or local density of states (LDOS).  In this
manuscript, we present a method by which the energy spectrum and the
eigenstates of a PQC can be obtained by solving Maxwell's equations in
higher dimensions for any PQC defined by the standard cut-and-project
construction, to which a generalization of Bloch's theorem applies.
In addition, we demonstrate how one can compute band structures with
defect states in the higher-dimensional superspace with no additional
computational cost.  As a proof of concept, these general ideas are
demonstrated for the simple case of one-dimensional quasicrystals,
which can also be solved by simple transfer-matrix techniques.
\end{abstract}

\maketitle

\section{Introduction}

We propose a computational method to solve for the spectra and
eigenstates of quasicrystalline electromagnetic structures by directly
solving a periodic eigenproblem in a higher-dimensional lattice.  Such
photonic quasicrystals (PQCs) have a number of unique properties
compared to ordinary periodic
structures~\cite{Chan98,Cheng99,Zoorob00:mse,Zoorob00:nat,DalNegro03,Wang03:jphys,Xie03,Notomi04,DellaVilla05,Feng05,Kim05,Lifshitz05,Vivas05,Wiersma05,DellaVilla06,Freedman06,Gauthier06:optcomm,Parker06,Zhang06,Zhang06:solcomm,Mnaymneh07},
especially in two or three dimensions where they can have greater
rotational symmetry and therefore offer some hope of complete photonic
band gaps with lower index
contrast~\cite{Zoorob00:nat,Kaliteevski01:jph,Zhang01,Hase02} than the
roughly $2$:$1$ contrast currently required for periodic
structures~\cite{Maldovan04}.  However, the study of two- and
three-dimensional photonic quasicrystals has been hampered by the
computational difficulty of modeling aperiodic structures, which has
previously required large ``supercell'' calculations that capture only
a portion of the infinite aperiodic lattice. Our method, in contrast,
captures the entire infinite aperiodic structure in a single
higher-dimensional unit cell, and we believe that this approach will
ultimately be much more computationally tractable for two- and
three-dimensional quasicrystals.  The idea that many quasicrystals can
be constructed by an irrational slice of a higher-dimensional lattice
is well known~\cite{Janot92,Stadnik99,Suck04}, and in fact is the most
common formulation of quasicrystals in two and three
dimensions~\cite{Wang03:jopt,Man05,Ledermann06}, but the possibility
of direct numerical calculations within the higher-dimensional space
seems to have been little explored outside of some tight-binding
calculations in quantum systems~\cite{Lange83,Lu87}.  As a proof of
concept, we demonstrate a first implementation of the technique
applied to one-dimensional quasicrystals, such as the well known
Fibonacci structure. Not only can we reproduce the spectrum from
transfer-matrix calculations, but we also show that the
higher-dimensional picture provides an interesting way to visualize
the eigenmodes and compute defect states in the infinite aperiodic
structure.

There have been several previous numerical approaches to simulating
quasicrystal structures in electromagnetism and quantum mechanics.  In
one dimension, a typical quasicrystal is an aperiodic sequence of two
or more materials, determined either by a slice of a
higher-dimensional lattice~\cite{Stadnik99} or by some ``string
concatenation'' rule~\cite{Janot92}.  In either case, efficient $2
\times 2$ transfer-matrix methods are available that allow one to
quickly compute the transmission spectra and density of states for
supercells consisting of many thousands of
layers~\cite{Godreche92,Huang01}.  Two- and three-dimensional
quasicrystals are almost always defined as an irrational slice (i.e.,
incommensurate Miller indices) of a higher-dimensional lattice; for
example, the famous Penrose tiling can be viewed as a two-dimensional
slice of a five-dimensional cubic lattice or of a four-dimensional
root lattice $A_4$~\cite{Suck04}.  In such cases, supercell
computations of a finite portion of the infinite aperiodic structure
(or a rational approximant thereof~\cite{Godreche92,Stadnik99})
require slower numerical methods, most commonly finite-difference
time-domain (FDTD)
simulations~\cite{Gauthier05,Kim05,Gauthier06:optcomm} or planewave
expansions~\cite{Kaliteevski00:jmo,DellaVilla06:ieee}.  Unfortunately,
these methods become very expensive for large supercells, nearly
prohibitively so for three-dimensional quasicrystals---there have been
experiments for 3D PQCs~\cite{Man05,Ledermann06}, but as yet few
theoretical predictions~\cite{Stuerer07,Zijlstra00}. With FDTD
methods, for example, the PQC local density of states is typically
integrated in Monte-Carlo fashion via random sources or initial
conditions~\cite{Wang03:jphys,DellaVilla05,Mnaymneh07}, but many
simulations are required to sample all possible modes in a large
supercell.  Also, the finite domain of a supercell becomes even more
significant in higher dimensions where a tractable supercell is
necessarily smaller, as there can be localized
states~\cite{Kim05,DellaVilla06,Gauthier06:optcomm,Mnaymneh07} whose
presence is dependent on the particular region of the PQC considered.
Our method of computing the spectrum directly in the
higher-dimensional unit cell, on the other hand, requires no supercell
to capture the infinite aperiodic structure---it uniformly samples (up
to a finite resolution) every possible supercell of the infinite
quasicrystal, rather than any particular subsection.  The influence of
finite-resolution on the convergence of the spectrum can be
systematically understood: one is not ``missing'' any part of the
quasicrystal, so much as resolving the entire quasicrystal with lower
resolution.

The structure of this paper is as follows: in \secref{cut} we review
the ``cut-and-project'' method for defining a PQC as a slice of a
higher-dimensional lattice, followed in \secref{computations} by a
description of our computational method in the higher-dimensional
lattice. There, we describe the extension of Maxwell's equations to
higher dimensions and also describe its solution in terms of a
higher-dimensional Bloch planewave expansion. As a proof of concept,
we present a sequence of one-dimensional examples in
\secref{results}. First, we compare results for a one-dimensional
``Fibonacci sequence'' with standard one-dimensional transfer-matrix
techniques. Second, as mentioned above, cut-and-project allows for a
straightforward way of studying defects in the the quasicrystal with
the same computational effort as the perfect PQC, and this is
demonstrated in the one-dimensional ``Fibonacci'' example. Finally, we
demonstrate the ease with which one can construct and explore
different quasicrystals by continuously varying the cut angle.

\section{Quasicrystals via cut-and-project}
\label{sec:cut}

Given a periodic lattice, any lower dimensional cross-section of that
lattice may be either periodic or quasi-periodic, depending upon the
angle of the cross-section.  For example, the periodic 2D
cross-sections of a 3D crystal are the lattice planes, defined in
crystallography by integer Miller indices.  If the Miller indices have
irrational ratios, on the other hand, the cross-section is aperiodic
but still has long-range order because of the underlying
higher-dimensional periodicity.  This is what is known as a
``cut-and-project'' method of defining a quasicrystalline structure:
as a slice of a periodic structure in a higher-dimensional
``superspace''~\cite{Janot92,Stadnik99}. (For a thorough discussion of
quasicrystals via cut-and-project, see \citeasnoun{Janot92}.)
Cut-and-project defines a specific class of quasicrystals;
equivalently, and more abstractly, cut-and-project corresponds to
structures whose Fourier transform has support spanned by a finite
number of reciprocal basis vectors (the projection of the reciprocal
lattice vectors from higher dimensions)~\cite{Janot92,Wang03:jopt}.
This class includes most commonly considered quasicrystals in two or
three dimensions, including the Penrose tiling~\cite{Suck04}, as well
as many one-dimensional quasicrystals including a version of the
Fibonacci structure.

\begin{figure}[t]
\includegraphics[width=0.4\textwidth]{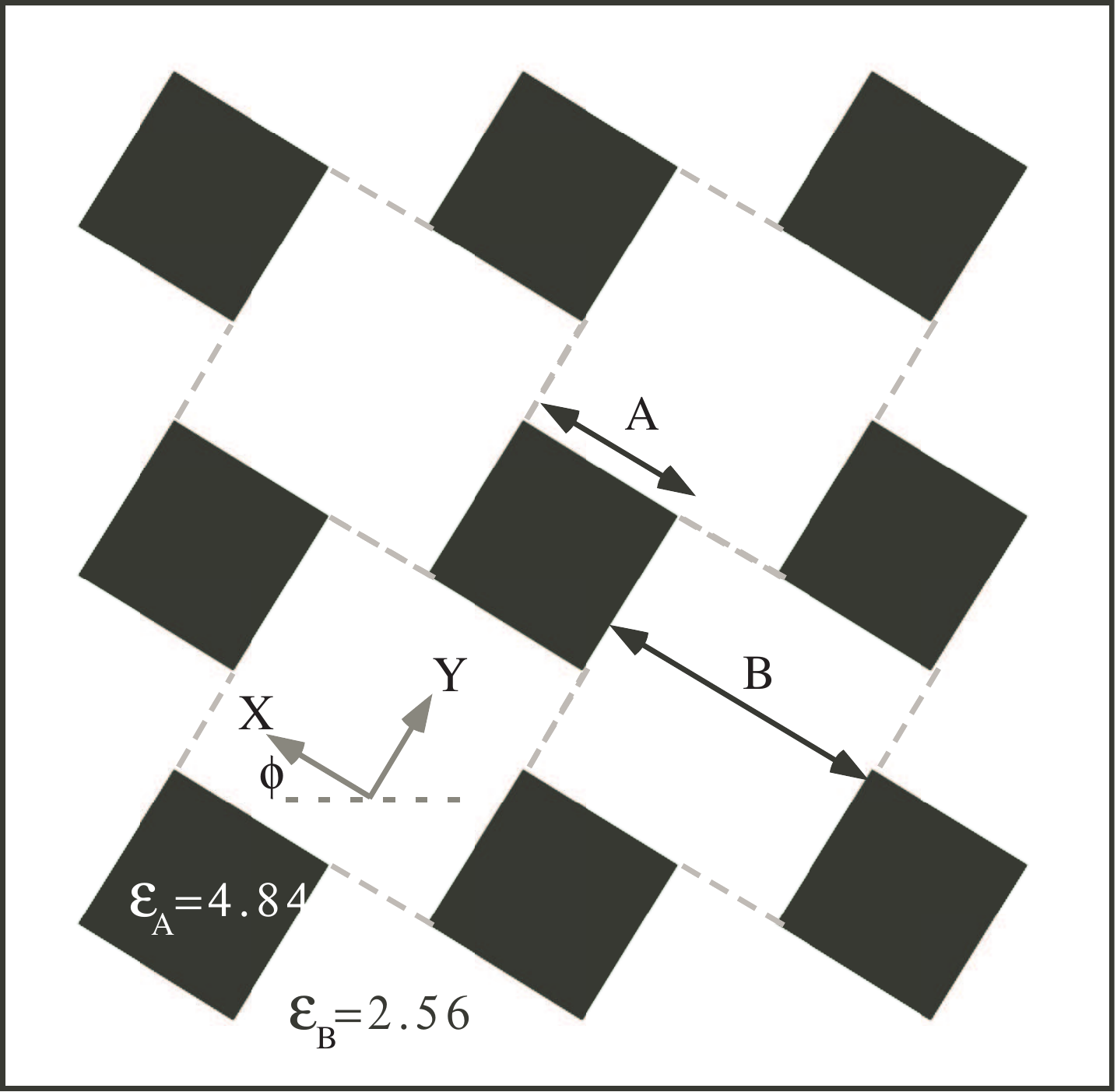}
\centering
\caption{Unit cell of the Fibonacci superspace dielectric. The
  physical dielectric is obtained by taking a slice at an angle
  $\tan{\phi}=\tau$. Black/white are the dielectric constants of the
  structure factor material and air, chosen to be $\varepsilon = 4.84$
  and $\varepsilon = 2.56$, respectively.}
\label{fig:geom}
\end{figure}

For example, consider the Fibonacci PQC in one dimension formed from
two materials $\varepsilon_A = 4.84$ and $\varepsilon_B = 2.56$ in
layers of thickness $A$ and $B$, respectively, similar to a recent
experimental structure~\cite{DalNegro03}. The Fibonacci structure $S$
is then defined by the limit $n\to\infty$ of the string-concatenation
rule $S_n = S_{n-2} S_{n-1}$ with starting strings $S_0=B$ and
$S_1=A$~\cite{DalNegro03}, generating a sequence $BABAABABAABA\cdots$.
In the case where $B/A$ is the golden ratio $\tau = (1+\sqrt{5})/2$,
exactly the same structure can be generated by a slice of a
two-dimensional lattice as depicted in \figref{geom}~\cite{Janot92}.
The slice is at an angle $\phi$ with an irrational slope $\tan\phi =
1/\tau$, and the unit cell of the 2D lattice is an $A \times A$ square
at an angle $\phi$ in a square lattice with period $(A+B)\sin\phi =
a$. Because the slope is irrational, the offset/intercept of the slice
is unimportant: any slice at an angle $\phi$ intercepts the unit cell
at infinitely many points, filling it densely.

For thickness ratios $B/A \neq \tau$, the Fibonacci structure cannot
be constructed by cut-and-project, and in general string-concatenation
rules can produce a different range of structures (such as the
Thue-Morse PQC~\cite{DalNegro04}) than cut-and-project.  This is partly a question
of definition---some authors reserve the term ``quasicrystal'' for
cut-and-project structures~\cite{Suck04}.  In any case, cut-and-project
includes a wide variety of aperiodic structures, including most of the
structures that have been proposed in two or three dimensions (where
they can be designed to have $n$-fold rotational symmetry for any
$n$), and are the class of quasicrystals that we consider in this
paper.

In general, let $d \leq 3$ be the number of physical dimensions of a
quasicrystal structure generated by a $d$-dimensional ``slice'' of an
$n$-dimensional periodic structure ($n>d$).  Denote this slice by $X$
(the physical space) with coordinates $\vec{x} \in \mathbb{R}^d$, and
denote the remaining $n-d$ coordinates by $\vec{y} \in
\mathbb{R}^{n-d}$ in the ``unphysical'' space $Y$ (so that the total
$n$-dimensional superspace is $Z = X \oplus Y$).  The primitive
lattice vectors $\vec{R}_i \in Z$ define the orientation of the
lattice with respect to the slice (rather than vice versa), with
corresponding primitive reciprocal vectors $\vec{G}_i$ defined by the
usual $\vec{R}_i \cdot \vec{G}_j = 2\pi
\delta_{ij}$~\cite{Janot92}. (The concept of an ``irrational slice''
is commonly used in the quasicrystal literature. However, a general
definition of what is meant by an ``irrational slice'' seems difficult
to find, and less evident in dimensions $d>2$. For a more precise
definition of ``irrational slice'' in general dimensions and a proof
that it is dense in the unit cell, see ~\secref{app}.)

The physical dielectric function $\varepsilon(\vec{x})$ is then
constructed by starting with a periodic dielectric function
$\varepsilon(\vec{x},\vec{y})$ in the superspace and evaluating it at
a fixed $\vec{y}$ (forming the slice).  Because an irrational slice is
dense in the unit cell of the superspace~\cite{Janot92}, it doesn't
matter what value of $\vec{y}$ one chooses, as discussed below (as
long as $\varepsilon$ is piecewise continuous). In principle, one
could define the unit cell of $\varepsilon$ in the superspace to be
any arbitrary $n$-dimensional function, but in practice it is common
to ``decorate'' the higher-dimension unit cell with extrusions of
familiar $d$-dimensional objects~\cite{Janot92,Suck04}.  More
precisely, ``cut-and-project'' commonly refers to constructions where
a set of lattice points within a finite window of the cut plane are
projected onto the cut plane, and this is equivalent to a simple cut
where objects at the lattice points are extruded in the $y$ direction
by the window width~\cite{Janot92}.  In particular, the extrusion
window is commonly an inverted projection (shadow) of the unit cell
onto the $y$ directions~\cite{Janot92}, although this is not the case
for the Fibonacci construction of \figref{geom}.

(Note that the higher-dimensional lattice need not be hypercubic.  For
example, the Penrose tiling can be expressed as a two-dimensional
slice of either a five-dimensional hypercubic lattice or of a
non-orthogonal four-dimensional root lattice $A_4$~\cite{Suck04}.  For computational
purposes, the lower the dimensionality the better.)

\section{Computations in Higher Dimensions}
\label{sec:computations}

Although the cut-and-project technique is a standard way to
\emph{define} the quasicrystal structure, previous computational
studies of photonic quasicrystals then proceeded to simulate the
resulting structure only in the projected ($d$-dimensional) physical
space.  Instead, it is possible to extend Maxwell's
equations into the periodic $n$-dimensional superspace, where Bloch's
theorem applies to simplify the computation.  By looking at only the
unit cell in $n$ dimensions one can capture the infinite
$d$-dimensional quasicrystal.  Our development of this technique was
inspired by earlier research on analogous electronic quasicrystals that applied a
tight-binding method in two dimensions to compute the spectrum of a
one-dimensional electronic quasicrystal~\cite{Lange83,Lu87}.

Let us start with Maxwell's equations in the physical space $X$ for
the quasicrystal $\varepsilon(\vec{x},\vec{y})$ at some fixed
$\vec{y}$ (that is, $\vec{y}$ is viewed as a parameter, not a
coordinate).  Maxwell's equations can be written as an eigenproblem
for the harmonic modes $\vec{H}(\vec{x},\vec{y}) e^{-i\omega
t}$~\cite{Joannopoulos95}, where again $\vec{y}$ appears as a
parameter.
\begin{equation}
\nabla_\vec{x} \times \frac{1}{\varepsilon(\vec{x},\vec{y})} \nabla_\vec{x} \times
\vec{H} = (\omega/c)^2 \vec{H} ,
\label{eq:eigenproblem}
\end{equation}
where $\nabla_\vec{x}\;\times$ denotes the curl with respect to the
$\vec{x}$ coordinates.  Assuming that the structure is
quasicrystalline, i.e. that $X$ is an irrational slice of the periodic
superspace $Z$, then $\omega$ should not depend upon $\vec{y}$~\cite{Lange83}.  The
reason is that $\vec{y}$ only determines the offset of the ``initial''
slice of the unit cell (for $\vec{x} = 0$), but as we reviewed above
the slice (considered in all copies of the unit cell) fills the unit
cell densely.  Therefore, any change of $\vec{y}$ can be undone, to
arbitrary accuracy, merely by offsetting $\vec{x}$ to a different copy
of the unit cell.  An offset of $\vec{x}$ doesn't change the
eigenvalues $\omega$, although of course it offsets the eigenfunctions
$\vec{H}$.

The fact that $\omega$ is independent of $\vec{y}$ allows us to
re-interpret \eqref{eigenproblem}, without actually changing anything:
we can think of $\vec{y}$ as a coordinate rather than a parameter, and
the operator on the left-hand side as an operator in $d$-dimensional
space.  Note that $\vec{H}$ is still a three-component vector field,
and $\nabla_\vec{x}\;\times$ is still the ordinary curl operator along
the $\vec{x}$ directions, so this is not so much a higher-dimensional
version of Maxwell's equations as an extension of the unmodified ordinary
Maxwell's equations into a higher-dimensional parameter space.  The $\vec{y}$
coordinate appears in the operator only through $\varepsilon$.
Because $\omega$ is independent of $\vec{y}$, i.e. it is just a number
rather than a function of the coordinates, the
equation~\eqnumref{eigenproblem} in higher dimensions is still an
eigenproblem, and its spectrum of eigenvalues $\omega$ is the same as
the spectrum of the $d$-dimensional quasicrystal, since the equations
are identical.  The physical solution is obtained by evaluating these
higher-dimensional solutions at a fixed $\vec{y}$, say $\vec{y}=0$
(where a different $\vec{y}$ merely corresponds to an offset in
$\vec{x}$ as described above).

For a real, positive $\varepsilon$, both the physical operator and the
extended operator in in \eqref{eigenproblem} are Hermitian and
positive semi-definite, leading to many important properties such as
real frequencies $\omega$~\cite{Joannopoulos95}.

\subsection{Bloch's theorem and numerics for quasicrystals}

Because the superspace eigenproblem is periodic, Bloch's theorem
applies: the eigenfunctions $\vec{H}(\vec{x},\vec{y})$ can be written
in the Bloch form $\vec{h}(\vec{z}) e^{i\vec{k}\cdot\vec{z}}$, where
$\vec{h}$ is a \emph{periodic} function defined by its values in the
unit cell, and $\vec{k}$ is the $n$-dimensional Bloch wavevector~\cite{Joannopoulos95}.

Here, $\vec{k}$ determines the phase relationship between $\vec{H}$ in
different unit cells of the superspace, but it does not have a simple
interpretation once the solution is projected into physical space.
The reason is that $\vec{h}$, viewed as a function of $\vec{x}$, is
again only quasiperiodic: translation in $\vec{x}$ ``wraps'' the slice
into a different portion of the unit cell, so both $\vec{h}$ and
$e^{i\vec{k}\cdot\vec{z}}$ change simultaneously and the latter phase
cannot be easily distinguished.  This prevents one from defining a
useful phase or group velocity of the PQC modes.

The key point is that Bloch's theorem reduces the eigenproblem to a
finite domain (the $n$-dimensional unit cell), rather than the
infinite domain required to describe the quasicrystal solutions in
physical space.  This means that standard numerical methods to find
the eigenvalues of differential operators are
immediately applicable.  For example, since the solution $\vec{h}$ is
periodic, one can apply a planewave expansion
method~\cite{Johnson2001:mpb} for $\vec{h}$:
\begin{equation}
\vec{h}(\vec{z}) = \sum_\vec{G} \tilde{\vec{h}}_\vec{G} e^{i\vec{G}\cdot\vec{z}},
\end{equation}
where the summation is over all $n$-dimensional reciprocal lattice
vectors $\vec{G}$.  Because the curl operations only refer to the
$\vec{x}$ coordinates, $\nabla_\vec{x} \times \vec{h}$ is replaced by
a summation over $\vec{G}_\vec{x} \times \tilde{\vec{h}}_\vec{G}$,
where $\vec{G}_\vec{x}$ denotes $\vec{G}$ projected into $X$.  The
resulting eigenproblem for the Fourier coefficients $\tilde{\vec{h}}$
(once they are truncated to some wavevector cutoff) can be computed
either by direct dense-matrix methods~\cite{Golub96} or, more efficiently, by
iterative methods exploiting fast Fourier transforms~\cite{Johnson2001:mpb}.  In the
present paper, we do the former, which is easy to implement as a proof
of concept, but for higher-dimensional computations an iterative
method will become necessary.

We should also remind the reader that there is a
constraint $\nabla_\vec{x}\cdot\vec{H} = 0$ on the eigenfunctions, in
order to exclude unphysical solutions with static magnetic charges.  In a
planewave method, this leads to a trivial constraint $(\vec{k}_\vec{x}
+ \vec{G}_\vec{x}) \cdot \tilde{\vec{h}} = 0$, again with $\vec{k}$
and $\vec{G}$ projected into $X$.

\subsection{The spectrum of the quasicrystal}
\label{sec:spectrum}

With a familiar eigenproblem arising from Bloch's theorem, such as
that of a periodic physical structure, the eigenvalues form a band
structure: discrete bands $\omega_n(\vec{k})$ that are continuous
functions of $\vec{k}$, with a finite number of bands in any given
frequency range~\cite{Kuchment01}.  For a finite-resolution
calculation, one obtains a finite number of these bands $\omega_n$ with
some accuracy that increases with resolution, but even at low
resolutions the basic structure of the low-frequency bands is readily
apparent.  The eigenvalues of the higher-dimensional quasicrystal
operator of \eqref{eigenproblem}, on the other hand, are quite
different.

The underlying mathematical reason for the discrete band structure of
a physical periodic structure is that the Bloch eigen-operator for a
periodic physical lattice,
$(\nabla+i\vec{k})\times\frac{1}{\varepsilon}(\nabla+i\vec{k})\,\times$,
is the inverse of a compact integral operator corresponding to the
Green's function, and hence the spectral theorem
applies~\cite{Gohberg00}. Among other things, this implies that the
eigenvalues at any given $\vec{k}$ for a finite unit cell form a
discrete increasing sequence, with a finite number of eigenvalues
below any finite $\omega$.  The same nice property does not hold for
the operator extended to $n$ dimensions, because along the $\vec{y}$
directions we have no derivatives, only a variation of the scalar
function $\varepsilon$.  Intuitively, this means that the fields can
oscillate very fast along the $\vec{y}$ directions without necessarily
increasing $\omega$, allowing one to have infinitely many
eigenfunctions in a finite bandwidth.  More mathematically, an
identity operator is not compact and does not satisfy the spectral
theorem~\cite{Gohberg00}, and since the operator of \eqref{eigenproblem}
is locally the identity along the $\vec{y}$ directions the same
conclusion applies. This means that, when the $\vec{y}$ direction is
included as a coordinate, it is possible to get an infinite number of
bands in a finite bandwidth at a fixed $\vec{k}$.

In fact, as we shall see below, this is precisely what happens, and
moreover it is what \emph{must} happen in order to reproduce the
well-known properties of quasicrystal spectra.  It has been shown that
quasicrystal spectra can exhibit a fractal structure~\cite{Janot92},
with infinitely many gaps (of decreasing size) in a finite bandwidth,
and such a structure could not arise from an ordinary band diagram
with a finite number of bands in a given bandwidth.  Of course, once
the unit cell is discretized for numerical computation, the number of
degrees of freedom and hence the number of eigenvalues is finite.
However, as the resolution is increased, not only do the maximum
frequency and the accuracy increase as for an ordinary computation,
but also the number of bands in a given bandwidth increases.  Thus, as
the resolution is increased, more and more of the fractal structure of
the spectrum is revealed.

\section{One-dimensional results}
\label{sec:results}

As a proof of concept implementation of cut-and-project, we construct
 a Fibonacci quasicrystal in \secref{fib-band} using the projection method
 described above, compute the band structure as a function of the
 projected wave-vector $k_{d}$ and compare to a transfer-matrix
 calculation of the same quasicrystal structure. We also demonstrate
 the field visualization enabled by the projection method,
 both in the superspace ($n$ dimensions) as well as in the
 physical space ($d$ dimensions). In \secref{defects}, we demonstrate
 how this method can accommodate systems with defects without
 additional computational costs. Finally, we explore several
 one-dimensional quasicrystal configurations in \secref{tilt} by
 varying the cut angle $\phi$.

\subsection{Fibonacci quasicrystal}
\label{sec:fib-band}

\begin{figure*}[t]
\centerline{\includegraphics[width=\textwidth]{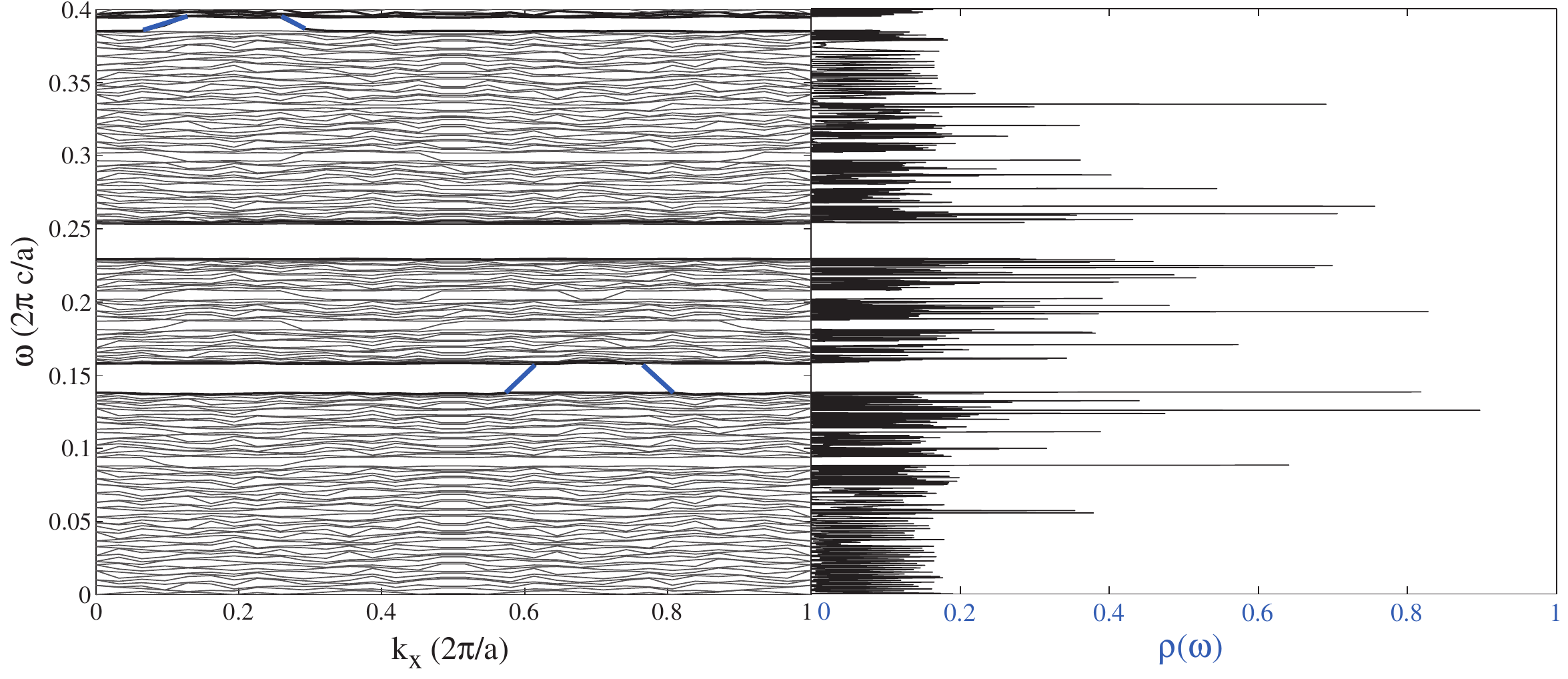}}
\caption{\emph{Left:} Frequency spectrum $\omega$ of the Fibonacci
  quasicrystal vs. ``wave-vector'' $k_x$. The blue lines indicate
  spurious states which arise due to finite resolution effects (see
  text). \emph{Right:} Corresponding density of states $\rho(\omega)$
  computed using a transfer-matrix technique with a supercell of $10^4$ layers. \emph{Inset}: Power
  distribution $\sim |H_z|^2$ (red/white = positive/zero) of a
  spurious state, the field profile of which oscillates at the Nyquist
  frequency.}
\label{fig:band-negro}
\end{figure*}

\begin{figure}[h]
\centerline{\includegraphics[width=0.49\textwidth]{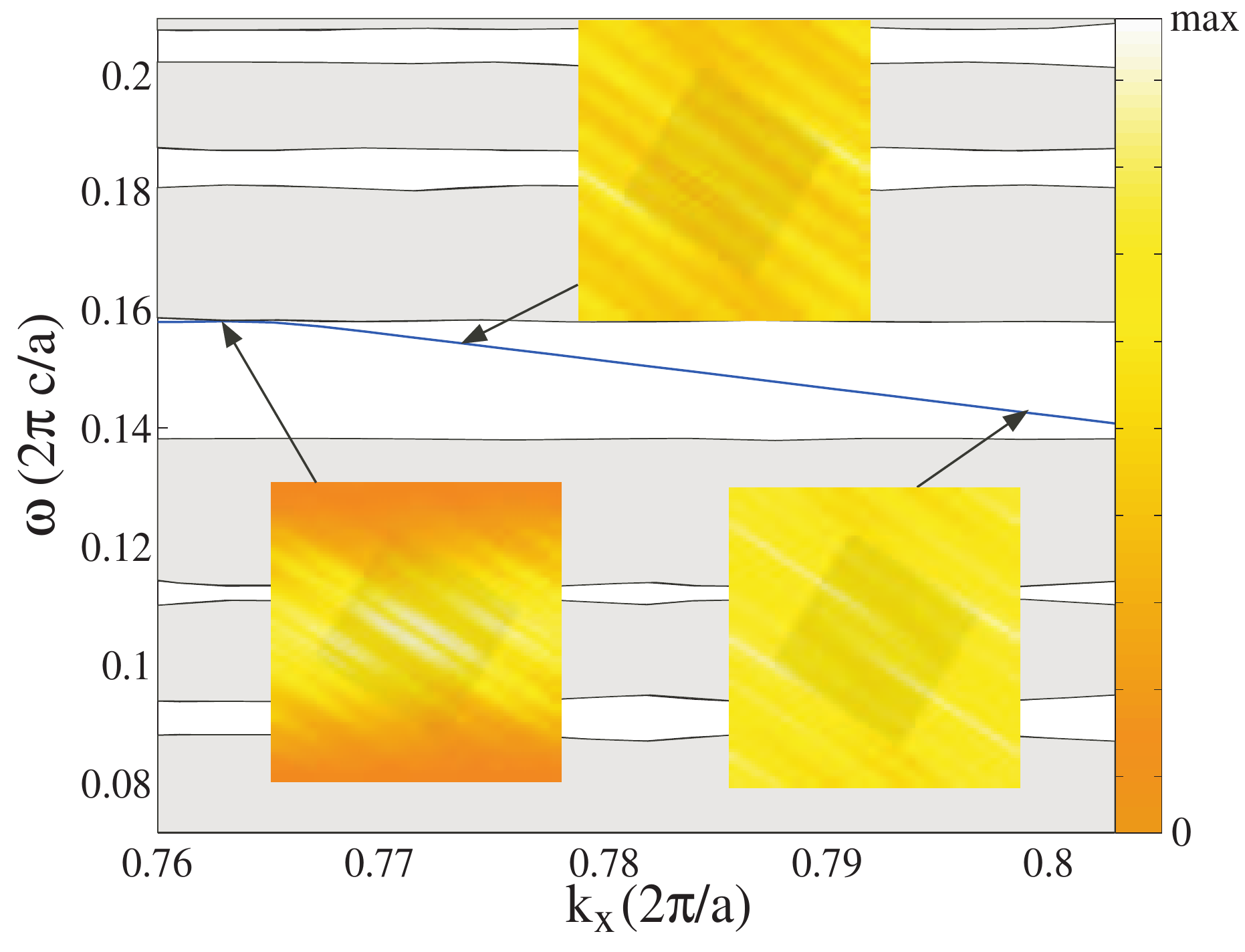}}
\caption{Enlarged view of the Fibonacci spectrum showing a gap with a
spurious band crossing it.  Insets show the magnetic field $|H_z|$ for
the spurious band at various $k_x$---the localization of this mode
around the $X$-parallel edges of the dielectric indicate that this is
a discretization artifact.}
\label{fig:spurious-mode}
\end{figure}

We solved \eqref{eigenproblem} numerically using a planewave expansion
in the unit cell of the 2D superspace, as described above, for the 1D
Fibonacci quasicrystal structure depicted in \figref{geom}.  The
resulting band diagram is shown in \figref{band-negro}(left), along
with a side-by-side comparison of the local density of states in
\figref{band-negro}(right) calculated using a transfer-matrix approach
with a supercell of $10^4$ layers~\cite{Li04}. The two calculations show
excellent agreement in the location of the gaps, except for one or two
easily-identified spurious bands inside some of the gaps, which are
discussed in further detail below.  The most important feature of
\figref{band-negro}(left) is the large number of bands even in the
finite bandwidth $\omega \in [0,0.4]$, with the number of bands
increasing proportional to the spatial resolution (planewave cutoff).
This is precisely the feature predicted abstractly above, in
\secref{spectrum}: at a low resolution, one sees only the largest
gaps, and at higher resolutions further details of the fractal
spectrum are revealed as more and more bands appear within a given
bandwidth, very different from calculations for periodic physical
media.  The important physical quantity is not so much the band
structure, since $\vec{k}$ has no simple physical meaning as discussed
previously, but rather the density of states formed by projecting the
band structure onto the $\omega$ axis.  In this density of states, the
small number of spurious bands within the gaps, which arise from the
discretization as discussed below, plays no significant role: the
density of states is dominated by the huge number of flat bands (going
to infinity as the resolution is increased), and the addition of one
or two spurious bands is negligible.

The ``spurious'' bands that appear within some of the band gaps of the
superspace calculation arise from the discretization of the dielectric
interfaces parallel to the slice direction.  Because the slice is at
an irrational angle, it will never align precisely with a uniform
grid, resulting in inevitable staircasing effects at the boundary.
With ordinary electromagnetic simulations, these staircasing effects
can degrade the accuracy~\cite{Farjadpour06}, but here the lack of derivatives
perpendicular to the slice allows spurious modes to appear along these
staircased edges (there is no frequency penalty to being localized
perpendicular to the slice).  Indeed, if one looks at the field
patterns for the spurious modes, one of which is shown in the inset of
\figref{band-negro}(right), one sees that the field intensity is
peaked along the slice-parallel dielectric interfaces.  Because they
are localized to these interfaces and therefore dominated by the
unphysical staircasing, the spurious modes behave quite differently
from the ``real'' solutions and are easily distinguished
qualitatively and quantitatively.  Most importantly, as the resolution
is increased, the number of spurious modes in a given gap does not
increase like all of the other bands, because the thickness of the
staircased interface region decreases proportional to the resolution.
This makes the gaps in the band structure obvious: here, they are the only
frequency ranges for which the number of eigenvalues does not increase
with resolution.  Equivalently, as noted above, the contribution of
the spurious bands to the density of states is asymptotically
negligible as resolution is increased.

\begin{figure}[t]
\centerline{\includegraphics[width=0.48\textwidth]{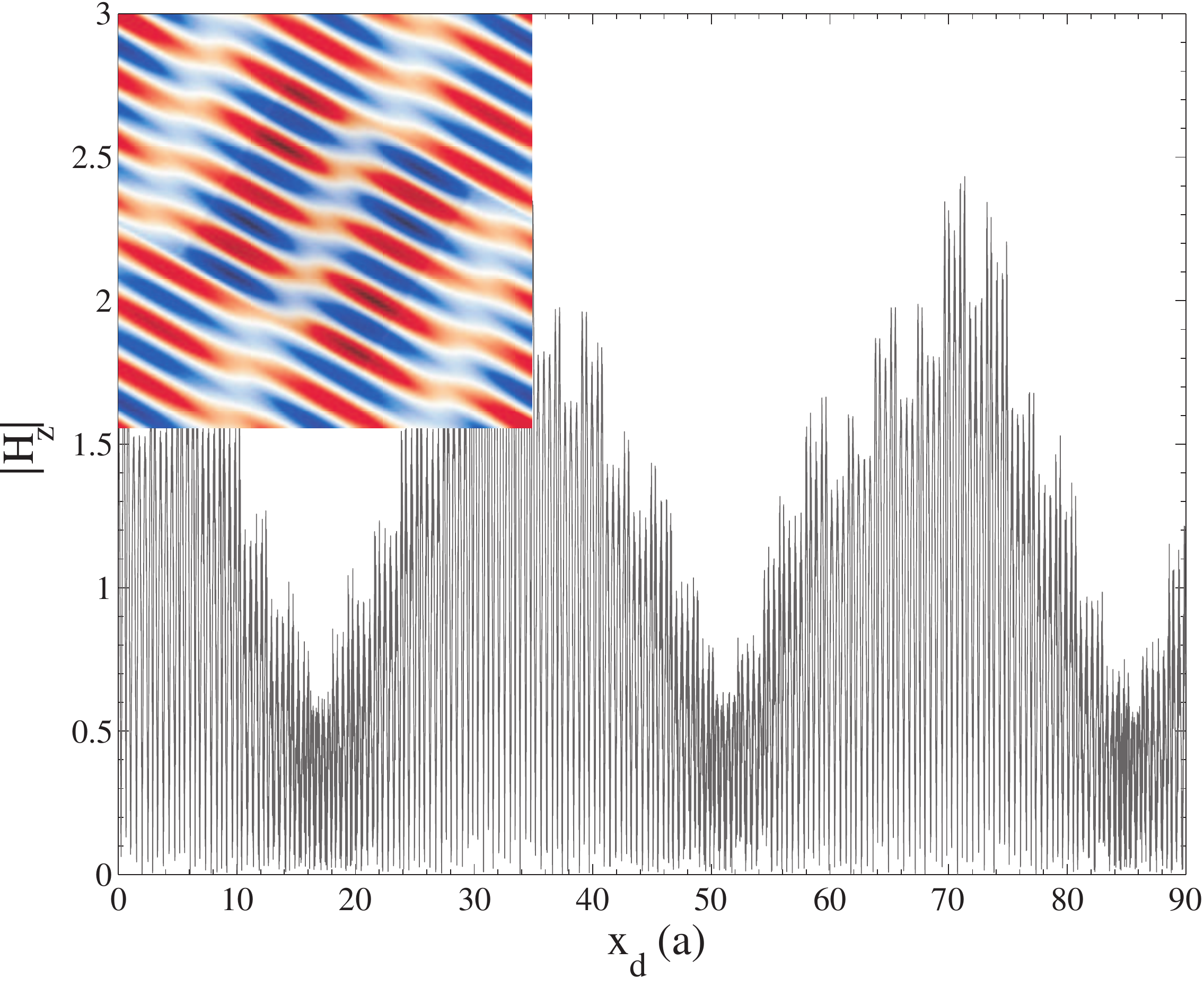}}
\caption{Plot of the magnetic field amplitude $|H_z|$ for a band-edge
state taken along a slice of the two-dimensional superspace (in the
$\phi$ direction). \emph{Inset:} Two-dimensional superspace field
profile (red/white/blue indicates positive/zero/negative
amplitude).}
\label{fig:extended}
\end{figure}

Computing the eigenmodes in the higher-dimensional superspace
immediately suggests a revealing visualization technique: instead of
plotting the quasiperiodic fields as a function of the physical
coordinates $x$ by taking a slice, plot them in the two-dimensional
superspace.  This has the advantage of revealing the entire infinite
aperiodic field pattern in a single finite plot~\cite{Lange83}.  One
such plot was already used above, to aid in understanding the spurious
modes localized at staircased interfaces.  A typical extended mode
profile is shown in \figref{extended}, plotted both as a function of
the physical coordinate $x$ for large supercell and also in the unit
cell of the superspace (inset).  In the inset superspace plot, one can
clearly see the predicted field oscillations perpendicular to the
slice plane, as well as a slower oscillation rate (inversely
proportional to the frequency) parallel to the slice.  In the plot
versus $x$, one can see the longer-range quasi-periodic structure that
arises from how the slice wraps around the unit cell in the
superspace.  The factor of three to four long-range variations in the
field amplitude are suggestive of the critically localized states
(power-law decay) that one expects to see in such
quasicrystals~\cite{Kohmoto83,Ostlund83,DalNegro03}.

\subsection{Defect modes}
\label{sec:defects}

\begin{figure}[t]
\includegraphics[width=0.4\textwidth]{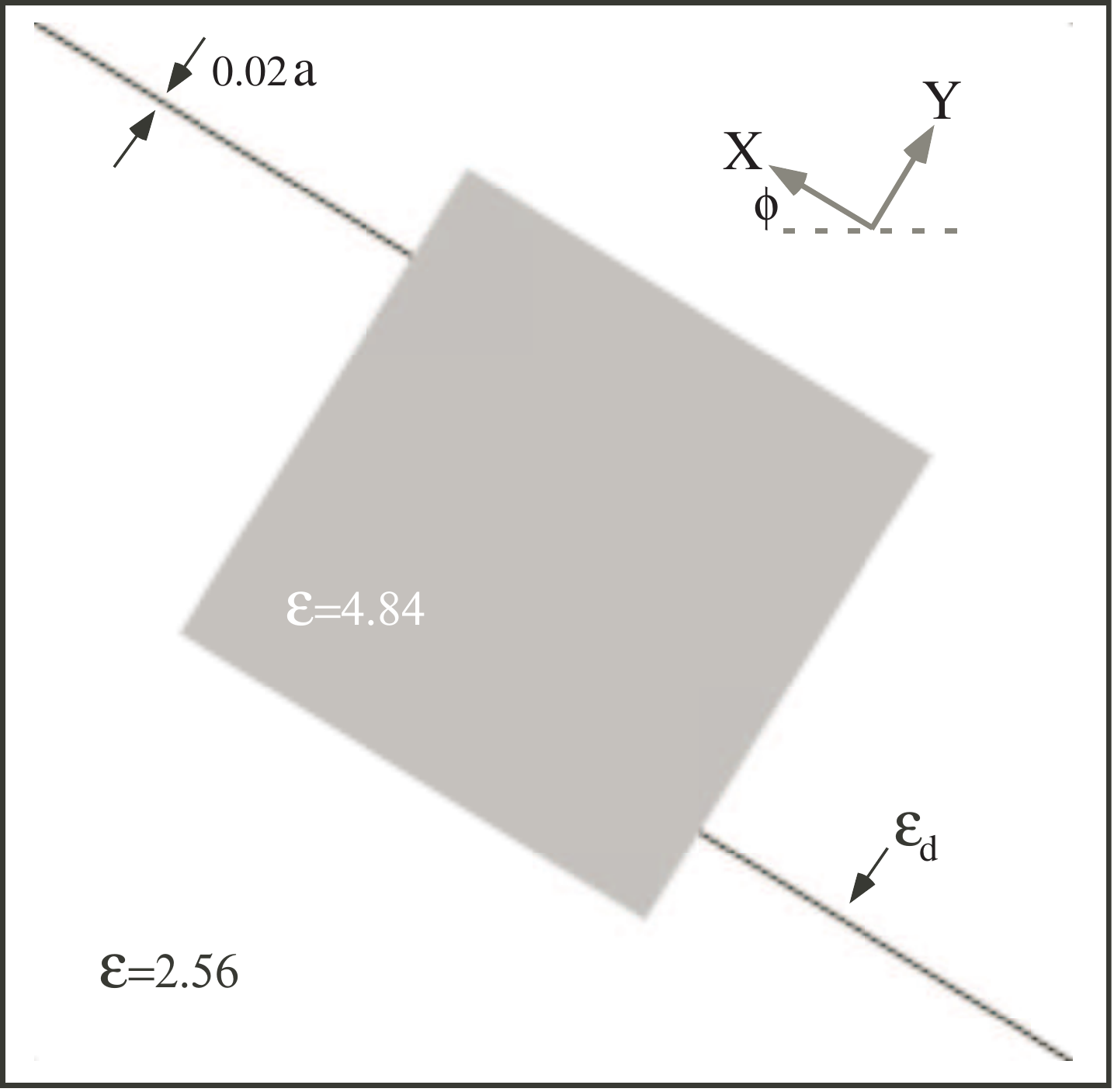}
\caption{Dielectric for the Fibonacci chain with $\varepsilon=2.56$
  (light blue), and a defect---an additional $\varepsilon=8.0$ layer,
  shown in gray.}
\label{fig:defect-eps}
\end{figure} 

Much of the interest in quasicrystal band gaps, similar to the
analogous case of band gaps in periodic structures, centers around the
possibility of localized states: by introducing a defect in the
structure, e.g. by changing the thickness of a single layer, one can
create exponentially localized states in the
gap~\cite{Cheng99,Bayindir01}.  In periodic systems, because such
defects break the periodicity, they necessitate a larger computational
cell, or supercell, that contains many unit cells.  In quasicrystal
systems, however, one can introduce a localized defect without
changing the higher-dimensional periodicity, and therefore compute
localized defect modes with the same superspace method and
computational cell.

Ideally, if one had infinite spatial resolution, a defect in the
crystal would be introduced as a very thin perturbation parallel to
the slice direction.  As the thickness of this perturbation goes to
zero, it intersects the physical slice at greater and greater
intervals in the physical space, corresponding to localized defects
that are separated by arbitrarily large distances.  In practice, of
course, the thickness of the perturbation is limited by the spatial
resolution, but one can still obtain defects that are very widely
separated---since the associated defect modes are exponentially
localized, the coupling between the defects is negligible.  In other
words, one effectively has a very large supercell calculation, but
expressed in only the unit cell of the higher-dimensional lattice.


\begin{figure}[t]
\includegraphics[width=0.47\textwidth]{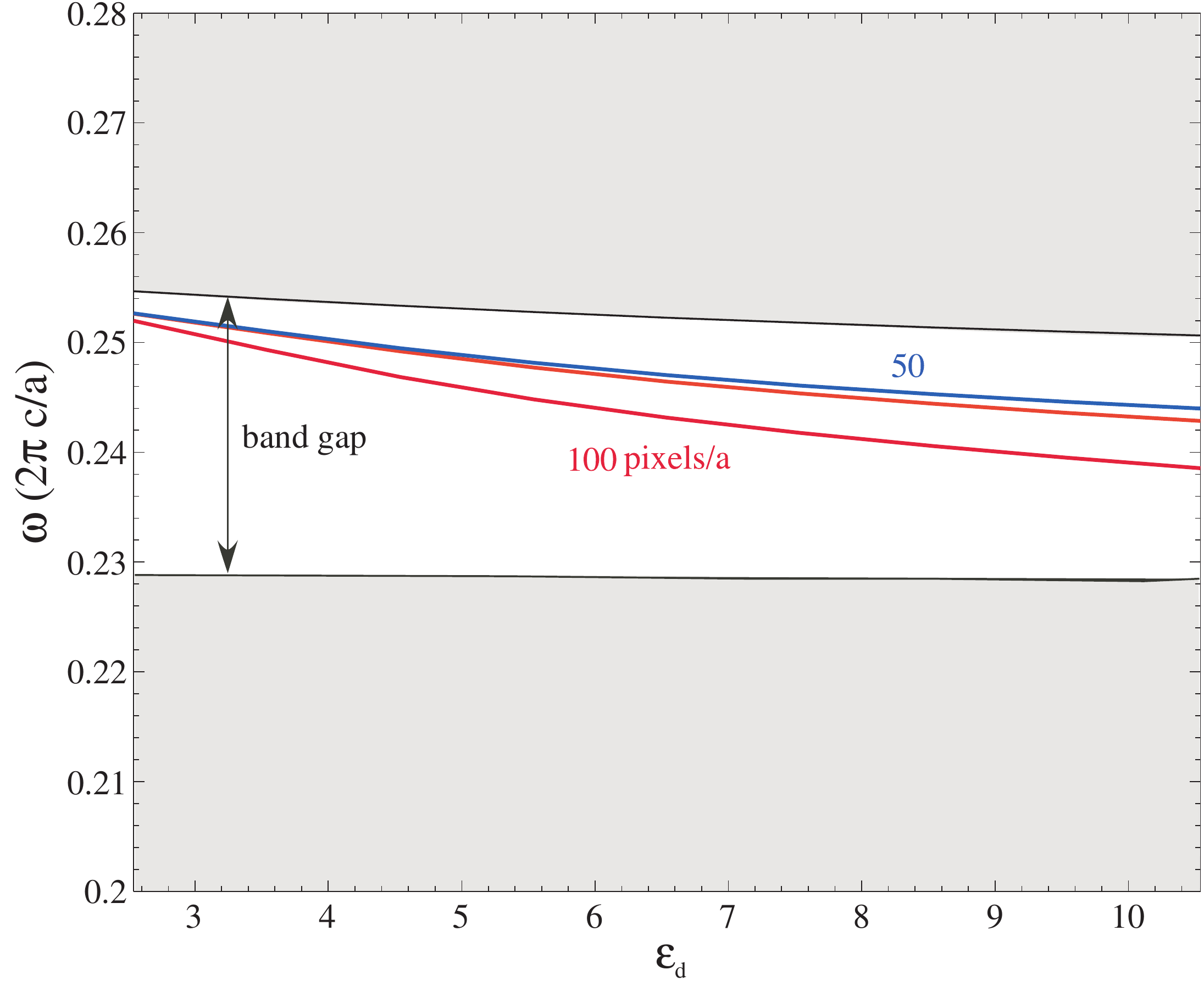}
\caption{Varying the defect epsilon for resolutions 50 (blue) and 100
  (red).  The thickness of the defect is fixed to 0.02 lattice
  constants.  The number of spurious modes increases with the
  resolution, the true defect state being the lowest of these modes.}
\label{fig:bands-epsilon}
\end{figure}

\begin{figure}[t]
\centerline{\includegraphics[width=0.48\textwidth]{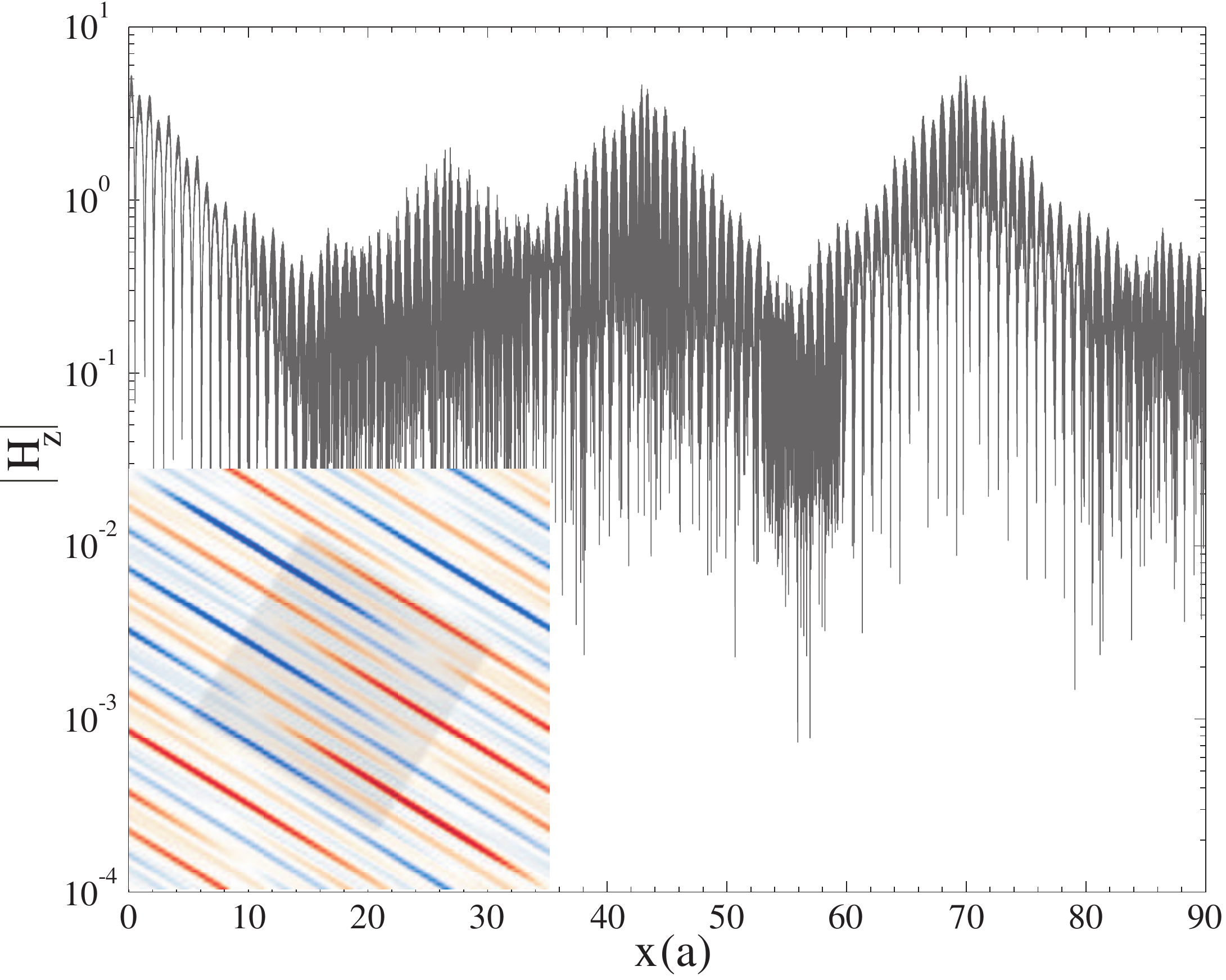}}
\includegraphics[width=0.48\textwidth]{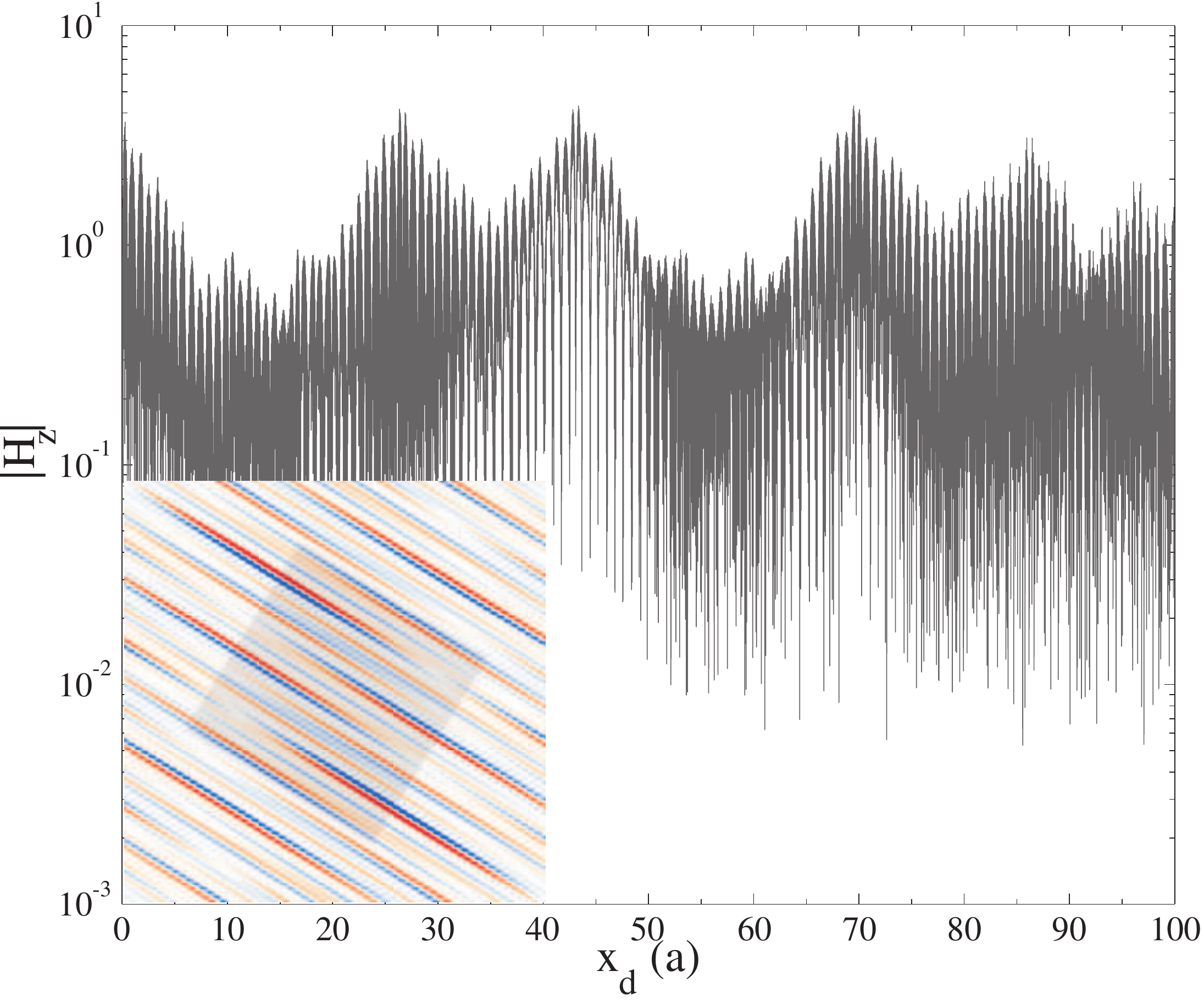}
\caption{Semi-log plots of the magnetic field magnitude $H_z$ for the
  lowest (top) and highest (bottom) defect state for the configuration
  shown in \figref{defect-eps}. \emph{Insets:} Two-dimensional
  superspace visualizations of the defect states. Note the
  additional node in the lower figure (corresponding to an unphysical
  oscillation).}
\label{fig:def}
\end{figure}

As an example, we changed an $\varepsilon=2.56$ layer to $\varepsilon
= \varepsilon_d$ at one place in the Fibonacci quasicrystal.  The
corresponding superspace dielectric function is shown in
\figref{defect-eps}, where the defect is introduced as a thin
($0.02a$) strip of $\varepsilon_d$ parallel to the slice direction. We
compute the band structure as a function of the defect dielectric
constant $\Delta\varepsilon = \varepsilon_d - 2.56$, varying it from
the normal dielectric $\varepsilon_d = 2.56$ up to $\varepsilon_d =
11$. The thickness of the defect in the unphysical direction was fixed
to be $\approx 0.02$. The resulting eigenvalues as a function of
$\Delta\varepsilon$ are shown in \figref{bands-epsilon} for two
different spatial resolutions of $50$ (blue) and $100$ (red)
pixels/$a$.  When the resolution is $50$ the defect is only one pixel
thick, the discretization effects might be expected to be large,
although the frequency is within about 2\% of the higher-resolution
calculation.  At the higher resolution, the frequency of the mode is
converging (it is within 0.3\% of a resolution-$200$ calculation, not
shown).  However, at the higher resolution there is a second, spurious
mode due to the finite thickness (2 pixels) of the defect layer---this
spurious mode is easily identified when the field is plotted
\figref{def}(bottom), because it has a sign oscillation perpendicular
to the slice (which would be disallowed if we could make the slice
infinitesimally thin).

The defect modes for the resolution $100$ are plotted in \figref{def}
for both the real and the spurious modes, versus the physical
coordinate ($x$) and also in the superspace unit cell (insets).  When
plotted versus the physical coordinate $x$ on a semilog scale, we see
that the modes are exponentially localized as expected.  The defect
mode appears at multiple $x$ values (every $\sim 20a$ on average)
because the defect has a finite thickness---the physical slice
intersects it infinitely many times (quasiperiodically), as discussed
above.  The spurious mode (bottom panel) is also exponentially
localized; it has a sign oscillation perpendicular to the slice
direction (inset) which causes it to have additional phase differences
between the different defects.

The advantages of the higher-dimensional (superspace) calculation over
a traditional supercell calculation are more tenuous for this sort of
defect calculation, because the exponential localization means that a
relatively small supercell can be employed.  On the other hand, this
is an illustration of the versatility of the superspace approach and
is a powerful tool for studying quasiperiodic structures with or without defects.

\subsection{Continuously varying the cut angle}
\label{sec:tilt}

\begin{figure}[ht]
\includegraphics[width=0.48\textwidth]{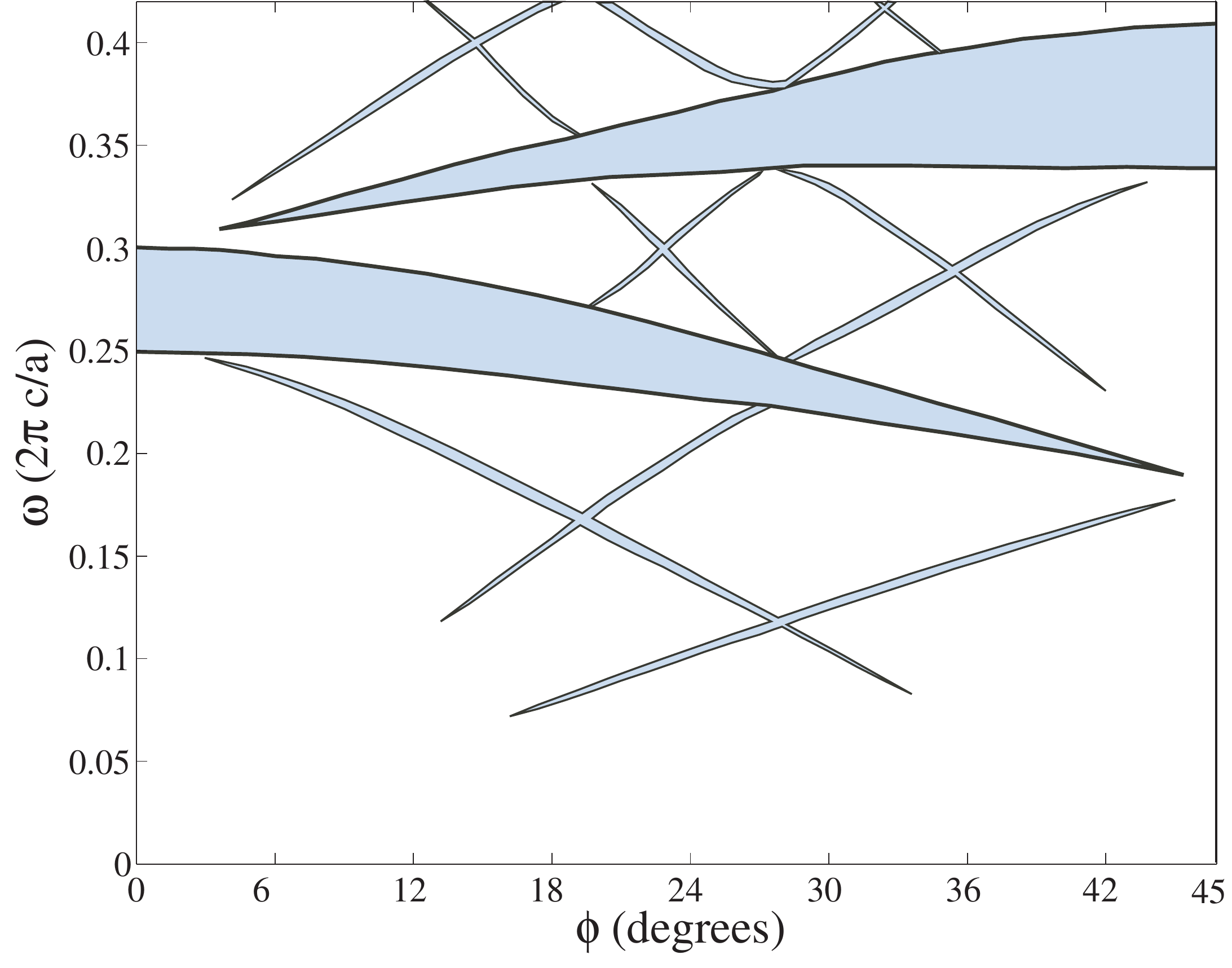}
\includegraphics[width=0.48\textwidth]{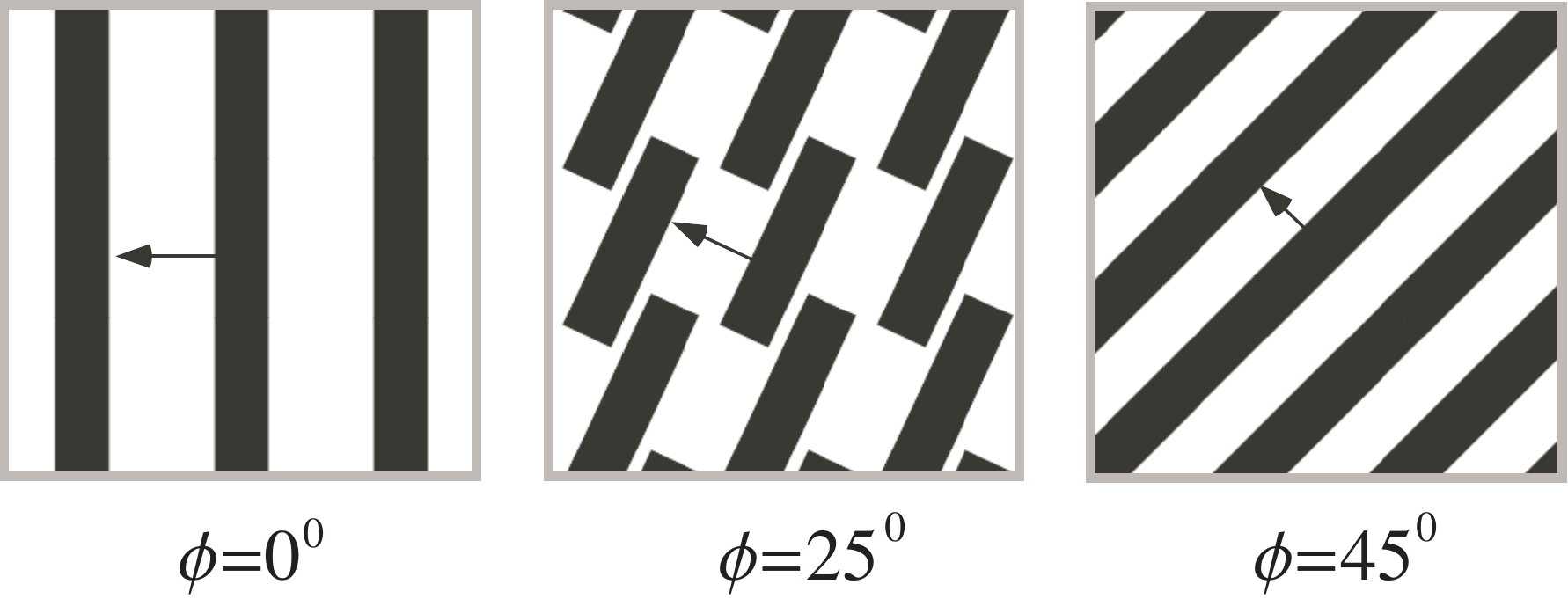}
\caption{Projected band structure vs. cut angle $\phi$, showing
different one-dimensional quasicrystal realizations.}
\label{fig:phi}
\end{figure}

The cut-and-project construction of quasicrystals provides a natural
way to parameterize a family of periodic and quasiperiodic structures,
via the cut angle $\phi$.  It is interesting to observe how the
spectrum and gaps then vary with $\phi$.

As $\phi$ is varied continuously from $0^\circ$ to $45^\circ$, the
structures vary from period $a$ to quasi-periodic lattices (for
$\tan\phi$ irrational) to long-period structures ($\tan\phi$ rational
with a large denominator) to a period $a\sqrt{2}$ crystal.  As we
change $\phi$, we rotate the objects in the unit cell, so that they
are always extruded along the $y$ direction with a length equal to the
projection of the unit cell onto $y$ [$a(\sin\phi + \cos\phi)$],
corresponding the usual cut-and-project construction~\cite{Janot92}.
In this case, the spectrum varies continuously with $\phi$, where the
rational $\tan \phi$ correspond to ``rational approximants'' of the
nearby irrational $\tan \phi$~\cite{Stadnik99,Wang03:jopt}.  For a
general unit cell with a rational $\tan \phi$, the physical spectrum
might depend on the slice offset $\vec{y}$ and hence different from
the total superspace spectrum, but this is not the case for dielectric
structures like the one here, which satisfy a ``closeness''
condition~\cite{Stadnik99} (the edges of the dielectric rods overlap
when projected onto the $Y$ direction).  This makes the structure
$\vec{y}$-independent even for rational slices~\cite{Stadnik99} The
resulting structures are shown in the bottom panel of \figref{phi} for
three values of $\phi$.

The corresponding photonic band gaps are shown in the top panel of
\figref{phi}, as a continuous function of $\phi$.  Only the largest
gaps are shown, of course, since we are unable to resolve the fractal
structure to arbitrary resolution.  As might be expected, there are
isolated large gaps at $\phi = 0^\circ$ and $\phi = 45^\circ$
corresponding to the simple $ABAB\cdots$ periodic structures at those
angles (with period $a$ and $a/\sqrt{2}$, respectively, the latter
resulting from two layers per unit cell).  The $\phi = 45^\circ$ gap
is at a higher frequency because of its shorter period, but
interestingly it is not continuously connected to the $\phi = 0^\circ$
gap.  The reason for this seems to be that the two gaps are dominated
by different superspace reciprocal lattice vectors: $(1,0)\cdot2\pi/a$
for $\phi=0^\circ$, and $(1,1)\cdot2\pi/a$ for $\phi=45^\circ$.  For
intermediate angles, a number of smaller gaps open and then close.  If
we were able to show the spectrum with higher resolution, we would
expect to see increasing numbers of these smaller gaps opening,
leading to the well-known fractal structure that arises e.g. for the
Fibonacci crystal.

This variation of gaps as a function of $\phi$ provides interesting
possibilities for band-gap engineering.  First, we see that we can get
large gaps that are close to one another in the spectrum, whereas in a
typical one-dimensional quarter-wave stack
the gaps are at integer multiples of a given frequency~\cite{Yeh88}.
Even more complex combinations of gaps may arise for higher
dimensional superspaces, since by including additional incommensurate
reciprocal lattice vectors one may generate additional nearby gaps.
Yet another interesting possibility would involve optimizing the layer
thicknesses as a function of $\phi$ so as to maximize the largest band
gap at every $\phi$ (or some other objective). The dielectric layers
chosen in \figref{phi} were of fixed thickness $0.37a$ (and variable
height equal to the projection of the unit cell along the $y$
direction as described above).

\section{Concluding Remarks}

We have presented a numerical approach to computing the spectra of
photonic quasicrystals by directly solving Maxwell's equations
extended to a periodic unit cell in higher dimensions, allowing us to
exploit Bloch's theorem and other attractive properties of
computations for periodic structures.  In doing so, we extended the
conceptual approach of cut-and-project techniques, which were
developed as a way to \emph{construct} quasicrystals, into a way to
\emph{simulate} quasicrystals.  Compared to traditional supercell
techniques, this allows us to capture the entire infinite aperiodic
quasicrystal in a single finite computational cell, albeit at only a
finite resolution.  In this way, the single convergence parameter of
spatial resolution replaces the combination of resolution and
supercell size in traditional calculations, in some sense uniformly
sampling the infinite quasicrystal.  The resulting computations,
applied to the test case of a Fibonacci quasicrystal, display the
unique features of quasicrystals in an unusual fashion, in terms of
higher-dimensional band structures and visualization techniques.  This
technique also allows defects and variation of cut angle (continuously
varying between periodic and aperiodic structures) in a
straightforward way.

In future work, we plan to apply this approach to modeling
higher-dimensional quasicrystal structures, where computing the
spectrum is currently more challenging using existing supercell
techniques.  To make a higher-dimensional superspace calculation
practical, one must use iterative eigensolver
methods~\cite{Johnson2001:mpb,bai00} rather than the simple
dense-matrix techniques employed for our test case.  Iterative
techniques are most efficient for computing a few eigenvalues at a
time, and so it will be useful to employ iterative methods designed to
compute ``interior'' eigenvalues~\cite{Johnson2001:mpb,bai00},
allowing one to search directly for large gaps without computing the
lower-lying modes.  Alternatively, numerical techniques have been
developed, based on filter-diagonalization methods, to directly
extract the spectrum of many eigenvalues without computing the
corresponding eigenvectors~\cite{Mandelshtam02}.

\section*{Appendix}
\label{sec:app}

In this appendix, we give an explicit derivation of the fact that an
``irrational'' slice densely fills the superspace unit cell, or rather
a definition of the necessary conditions to be an ``irrational'' slice.
These concepts are widely used in the quasicrystal literature, but a
precise definition seems hard to find (one commonly requires that all
of the Miller indices have incommensurate ratios, but this condition
is stronger than necessary).

Without loss of generality, we can consider the unit cell in the
superspace $Z=\mathbb{R}^n$ to be the unit cube (related to any
lattice by an affine transformation) with lattice vectors along the
coordinate directions.  The physical slice $X$ is $d$-dimensional, and
it will be convenient to write the coordinates of a vector $\vec{z}$
as $\vec{z} = (s_1,\ldots,s_d,t_1,\ldots,t_{n-d}) =
(\vec{s},\vec{t})$.  By taking every coordinate modulo $1$, we can map
$X$ to a set $\bar{X}$ consisting of $X$'s intersection with each unit cell.
We wish to show necessary and sufficient conditions for $\bar{X}$ to
densely fill the unit cell.  

\begin{figure}[h]
\includegraphics[width=0.47\textwidth]{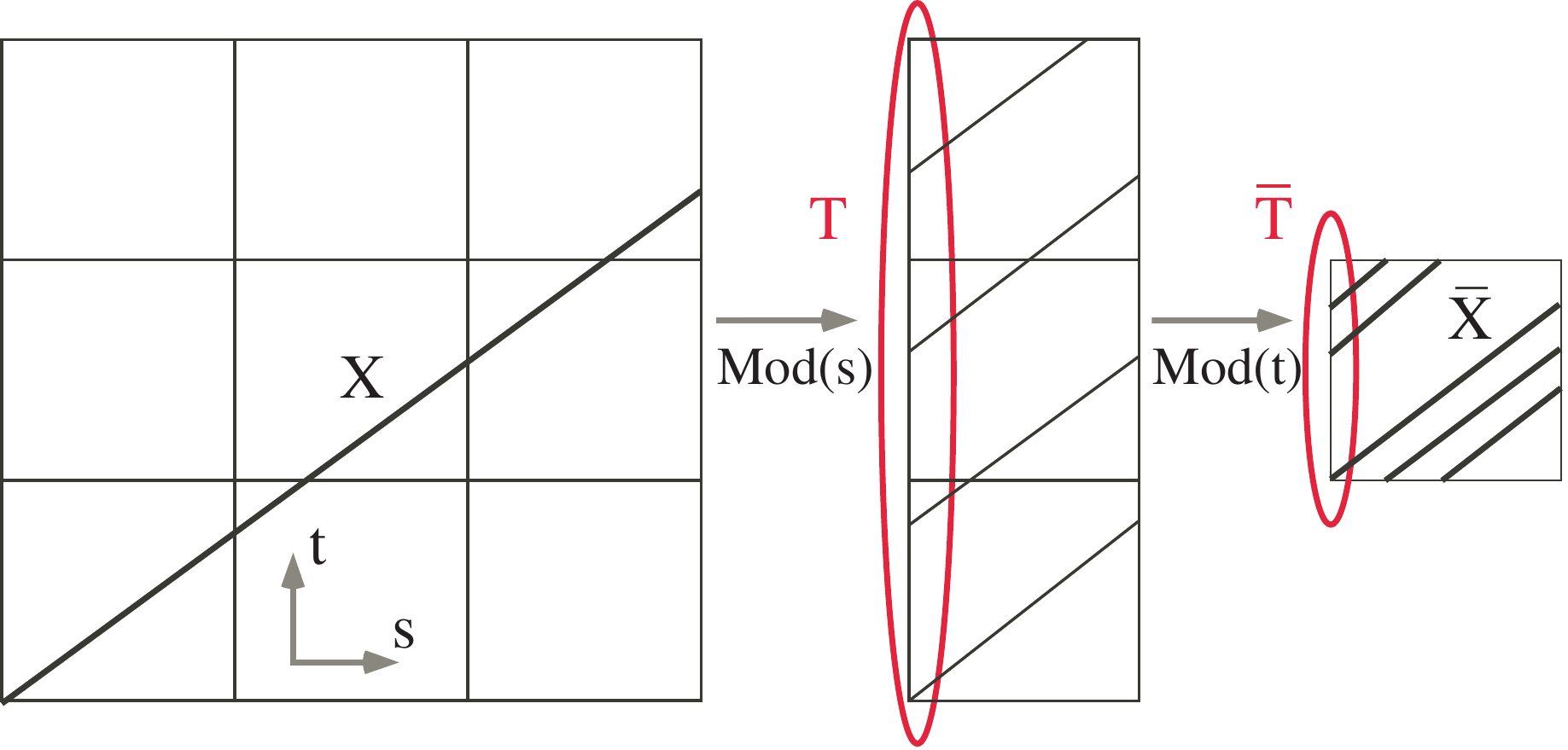}
\centering
\caption{An example two-dimensional superspace slice $X$ (left) and
the projected slice modulo~1 into the unit cell $\bar{X}$ in the $s$
(middle) and $t$ directions (right), along with the intersections $T$
(middle, red) and $\bar{T}$ (right, red) of $\bar{X}$ with the $s=0$
hyperplane.}
\end{figure}

Assuming that the slice is not orthogonal to any of the coordinate
axes (as otherwise it would clearly not densely fill the unit cell), we
can parameterize the points $\vec{z}$ of $X$ so that the last $n-d$
coordinates $(t_1, \ldots ,t_{n-d})$ are written as a linear function
$\vec{t}(s_1,\ldots s_d) \equiv \vec{t}(\vec{s})$ of the first $d$
coordinates.

Consider the set $T$ in $\mathbb{R}^{n-d}$ formed by the
$\vec{t}(\vec{s})$ coordinates of $X$ when the components of $\vec{s}$
take on integer values.  This is a subset of $X$, and the
corresponding set $\bar{T}$ formed by taking $\vec{t} \in T$ modulo~1
is a subset of $\bar{X}$.  The key fact is that $\bar{X}$ is dense in
the $n$-dimensional unit cell if and only if $\bar{T}$ is dense in the
$(n-d)$-dimensional unit cell, and this is the case that we will
analyze.  This equivalence follows from the fact that $\bar{X}$ is
simply $\bar{T}$ translated continuously along the slice directions
(every point in $\bar{X}$ is related to a point in $\bar{T}$ by a
simple projection). The set $T$ is a lattice in $\mathbb{R}^{n-d}$
consisting of all integer linear combinations of the basis vectors
$\vec{t}_k = \vec{t}(s_j = \delta_{jk})$, since $\vec{t}(\vec{s})$ is
a linear function.

For each basis vector $\vec{t}_k$, it is a well-known
fact~\cite{Ott02} that if it consists of $m$ incommensurate irrational
components, the set of integer multiples $\ell \vec{t}_k$ modulo~1
will densely fill an $m$-dimensional slice of the unit cell.  More
precisely, write $\vec{t}_k = \sum_{j=1\ldots m_k} \alpha_k^j
\vec{b}_k^j + \vec{q}_k$, where the $\vec{b}_k^j$ and $\vec{q}_k$ have
purely rational components and the $\{ \alpha_j \}$ are incommensurate
irrational numbers, and $m_k$ is therefore the number of
incommensurate irrational components of $\vec{t}_k$.  Then the set of
integer multiples of $\vec{t}_k$ modulo~1 densely fills an
$m_k$-dimensional slice of the unit cell of $\mathbb{R}^{n-d}$. The
basis vectors of this slice are precisely the vectors $\vec{b}_k^j$,
which are rational and therefore commensurate with the basis vectors
of $\mathbb{R}^{n-d}$, while the vector $\vec{q}_k$ is simply a
rational shift.  This slice therefore cuts the unit cell of
$\mathbb{R}^{n-d}$ a finite number of times.

The set $\bar{T}$ is then obtained as the direct sum of these dense
slices for all $n-d$ vectors $\vec{t}_k$.  This is then dense if and
only if the set of vectors $\{ \vec{b}_k^j \}^{j=1\ldots
m_k}_{k=1\ldots d}$ spans $\mathbb{R}^{n-d}$.  In other words, an
``irrational slice,'' which densely fills the unit cell, is one in
which there are $n-d$ independent incommensurate slice components as
defined above.

\end{document}